\documentclass{aa}  

\usepackage{graphicx}
\usepackage{natbib}

\usepackage{booktabs}

\graphicspath{{./Pictures/}} 

\usepackage{txfonts}

\begin{document}

   \title{The GAPS Programme with HARPS-N at TNG}

   \subtitle{XV. A substellar companion around a K giant star identified with quasi-simultaneous HARPS-N and GIANO measurements \thanks{Based on observations collected at the Italian Telescopio Nazionale Galileo (TNG), operated on the island of La Palma by the Fundación
Galileo Galilei of the INAF (Istituto Nazionale di Astrofisica) at the
Spanish Observatorio del Roque de los Muchachos of the Instituto
de Astrofísica de Canarias, in the frame of the programme Global
Architecture of Planetary Systems (GAPS)}}

   \author{E. Gonz\'alez - \'Alvarez 
          \inst{1,2}
          \and  L. Affer
          \inst{1}
          \and  G. Micela
          \inst{1}
          \and  J. Maldonado
          \inst{1}
          \and  I. Carleo
          \inst{3,4}
          \and M. Damasso
          \inst{4,5}
          \and V. D'Orazi
          \inst{3}
          \and {A. F. Lanza}
          \inst{6}  
          \and K. Biazzo
          \inst{6}
          \and  E. Poretti
          \inst{7}
          \and R. Gratton
          \inst{3}
          \and  A. Sozzetti
          \inst{5}
          \and S. Desidera
          \inst{3}
          \and  N. Sanna
          \inst{8}
          \and A. Harutyunyan
          \inst{9}
          \and F. Massi
          \inst{8}
          \and E. Oliva
          \inst {8}     
          \and R. Claudi
          \inst {3}
          \and R. Cosentino
          \inst {9}
          \and E. Covino
          \inst {10}
          \and A. Maggio
          \inst {1}
          \and S. Masiero
          \inst {1}
          \and E. Molinari
          \inst {9,11}
          \and I. Pagano
          \inst {6}
          \and G. Piotto
          \inst {3,4}
          \and R. Smareglia
          \inst {12}
          \and S. Benatti
          \inst {3}
          \and A. S. Bonomo
          \inst {5}
          \and F. Borsa
          \inst {7}
          \and M. Esposito
          \inst {10}
          \and P. Giacobbe
          \inst {5}
          \and L. Malavolta
          \inst {3,4}
          \and A. Martinez-Fiorenzano
          \inst {9}
          \and V. Nascimbeni
          \inst {3,4}
          \and M. Pedani
          \inst {9}
          \and M. Rainer
          \inst {7}
          \and G. Scandariato
          \inst {6}          
          }
         
             \institute{INAF-Osservatorio Astronomico di Palermo,
              Piazza Parlamento 1, 90134 Palermo, Italia
              \and Dipartimento di Fisica e Chimica - Universit\`a degli Studi di Palermo, via Archirafi 36, 90123 Palermo, Italia
               \and INAF-Osservatorio Astronomico di Padua, Vicolo dell’Osservatorio 5, 35122 Padova, Italia
               \and Dipartimento di Fisica e Astronomia G. Galilei, Universit\`a di Padova, Vicolo dell’Osservatorio 2, 35122 Padova, Italia
               \and INAF-Osservatorio Astrofisico di Torino, via Osservatorio 20, 10025 Pino Torinese, Italia
               \and INAF-Osservatorio Astrofisico di Catania, via S. Sofia 78, 95123 Catania, Italia
               \and INAF-Osservatorio Astronomico di Brera, via E. Bianchi 46, 23807 Merate, Italia
               \and INAF-Osservatorio Astronomico di Arcetri, Largo Enrico Fermi, 5, 50125 Firenze, Italia
               \and Fundaci\'on Galileo Galilei-INAF, Rambla Jos\'e Ana Fernandez P\'erez 7, 38712 Bre\~na Baja, TF, Espa\~na
               \and INAF-Osservatorio Astronomico di Capodimonte, Salita Moiariello 16, 80131 Napoli, Italia
               \and INAF-IASF Milano, via Bassini 15, 20133 Milano, Italia  
               \and INAF-Osservatorio Astronomico di Trieste, via Tiepolo 11, 34143 Trieste, Italia
				}

   \offprints{E. Gonz\'alez - \'Alvarez \\ \email{egonzalez@astropa.inaf.it}}

 
  \abstract
   {Identification of planetary companions of giant stars is made difficult because of the astrophysical noise, that may produce radial velocity variations similar to those induced by a companion. On the other hand any stellar signal is wavelength dependent, while signals due to a companion are achromatic.}
   {Our goal is to determine the origin of the Doppler periodic variations observed in the thick disk K giant star TYC 4282-605-1 by HARPS-N at the Telescopio Nazionale Galileo (TNG) and verify if they can be due to the presence of a substellar companion.}
   {Several methods have been used to exclude the stellar origin of the observed signal including detailed analysis of activity indicators and bisector and the analysis of the photometric light curve. Finally we have conducted an observational campaign to monitor the near infrared (NIR) radial velocity with GIANO at the TNG in order to verify whether the NIR amplitude variations are comparable with those observed in the visible.}
   {Both optical and NIR radial velocities show consistent variations with a period at 101 days and similar amplitude, pointing to the presence of a companion orbiting the target. The main orbital properties obtained for our giant star with a derived mass of $M=0.97 \pm 0.03$ $M_{\odot}$ are $M_P\sin i = 10.78\pm 0.12 \; M_{\rm J}$; $P=101.54 \pm 0.05$ days; $e = 0.28 \pm 0.01$ and $a=0.422 \pm 0.009$ AU. The chemical analysis shows a significant enrichment in the abundance of Na~{\sc i}, Mg~{\sc i}, Al~{\sc i} and Si~{\sc i} while the rest of analyzed elements are consistent with the solar value demonstrating that the chemical composition corresponds with an old K giant (age = 10.1 Gyr) belonging to local thick disk.}
   {We conclude that the substellar companion hypothesis for this K giant is the best explanation for the observed periodic radial velocity variation. This study also shows the high potential of multi-wavelength radial velocity observations for the validation of planet candidates.}


\maketitle
%

\section{Introduction}
\label{Introduction}

Several physical mechanisms related to the host star characteristics may mimic the presence of a planet. They include granulation and activity phenomena such as spots, plages and even cycles as well as radial and non-radial pulsations \citep{2000AJ....120..979H,2011A&A...527A..82D}. These phenomena may occur on several time scales and produce radial velocity (RV) variations hardly distinguishable from those induced by a low-mass companion.

The K giant stars usually have masses in the 0.2-10 M$_{\odot}$ mass range and radii between 2 and 100 R$_{\odot}$. Their precise evolutionary stage is difficult to determine because the evolutionary tracks of the stars belonging to the red giant branch, red clump (for solar metallicity) and asymptotic giant branch fall close and can intersect with each other in the H-R diagram \citep{2013ARA&A..51..353C}.

The variability of these objects can be quite complex.
Luminous red-giants are known to exhibit regular and semiregular
light curves, with periods of tens and hundreds of days, and
additional long-term modulations not yet fully understood \citep{2004ApJ...604..800W}. Several mechanisms have been suggested, such as  rotational modulation, additional pressure or gravity pulsation modes, mass loss effects induced by companions, amongst others. The space mission {\it CoRoT} disclosed the huge asteroseismic potential of G-K giants: the regular patterns due to solar-like oscillations and the progressive shift of frequency of maximum oscillation power were both observed \citep{2009Natur.459..398D}. Further detailed investigations of the pulsational properties allowed asteroseismologists to investigate the internal stellar structure \citep[see][for reviews]{2013ARA&A..51..353C,2016arXiv160907487H}.

Surface features and non-radial pulsations are expected to produce changes in the shape of the spectral line profile which can be misinterpreted with velocity shifts. These changes can be measured using spectral line bisectors that are correlated with RV in case of distortion of the lines. The measurements of the bisector require very high resolution spectra and a stable instrument. The absence of variations in the spectral line shape is a necessary condition but it is not sufficient to prove the existence of planetary companion.

Photometric observations are useful to understand the long-period RV variations as they constitute an independent measure of the time scales of rotation and pulsation periods. A photometric variability on the same time scale of the RV variations would immediately exclude the planetary companion hypothesis. Photometric analysis, together with the monitoring of the Ca II H and K lines (measured using the same spectra as to derive RVs) and bisector analysis are very helpful to reject the companion hypothesis and should always be used when possible.

As mentioned before, the pulsations in K giants are excited
by p-modes spanning a large interval of periods. The amplitudes
of the RV variations are expected to show an atmospheric 
gradient as it happens for Cepheids \citep[see Fig.~7 in][]{2017A&A...597A..73N}. The processes of stellar origin are chromatic, therefore measurements of the RV in more than one spectral band may be helpful to discriminate the origin (stellar or keplerian) of the observed variations. Pulsations cause light variations, and in this case the photometric amplitudes are also different at visible and infrared wavelengths \citep{2001PASP..113..983P}.

A similar behaviour is expected in the case of activity, where the variations are due to the difference of temperature between the unperturbed photosphere and the spot. On the contrary the planetary signal is independent from the spectral band, therefore the comparison of the RV amplitude variations in the two bands may provide an effective way to discriminate stellar and keplerian variations.

In this paper we present the analysis of the RV variations of the giant star TYC 4282-605-1 observed with HARPS-N at the TNG. These observations were obtained within the Global Architecture of Planetary Systems \citep[GAPS, see][]{2013A&A...554A..28C} observing programme. The GAPS programme started its operations in August 2012 taking advantage of the high performance of the HARPS-N spectrograph \citep{2012SPIE.8446E..1VC}, mounted at the Italian telescope TNG in La Palma, Canary Islands.  

The observations showed a periodic signal in RV that could be attributed to a planet. Since we cannot exclude a possible alternative explanation, we acquired additional data in the near infrared (NIR) with the GIANO spectrograph \citep{2006SPIE.6269E..19O} at the TNG, to obtain quasi-simultaneous observations in optical and NIR to discriminate between a stellar and keplerian origin of the periodicity observed in the optical band. 

In Sect. \ref{Stellar properties} we derive the stellar parameters and discuss the chemical analysis. Section \ref{The HARPS-N observations} describes the observations with HARPS-N, the RV analysis of the collected spectra, the effects of the stellar contribution to RV variation through the study of the bisector as asymmetry indicator and finally the chromospheric emission from Ca II H and K lines. Section \ref{Photometric analysis} describes the photometric analysis. Our methods of extracting the RV in NIR are given in Sect. \ref{The GIANO near-IR observations} together with observations, data reduction of the GIANO spectrograph and the consistency between optical and NIR data. Finally, the conclusions are presented in Sect. \ref{Summary and conclusions}.

\begin{table}[h]
\centering
\caption{Stellar parameters of TYC 4282-605-1 from literature.}
\label{stellar_parameters_1}
\begin{tabular}{c c}
\hline
\hline
\noalign{\smallskip}
Parameters  & Value \\
\noalign{\smallskip}	
\hline	
\noalign{\smallskip}
  
$\rm \alpha$ (J2000) &   22 55 29.2581$^{(a)}$	\\ 
$\rm \delta$ (J2000) &   +62 14 20.849$^{(a)}$\\
B [mag] & $12.105 \pm 0.01^{(a)}$\\
V [mag]  & $10.58 \pm 0.01^{(a)}$\\
B-V [mag] & $1.52 \pm 0.01$\\
J [mag] & $7.667 \pm 0.023^{(b)}$ \\ 
H [mag] & $6.963 \pm 0.033^{(b)}$\\
K [mag] & $6.778 \pm 0.021^{(b)}$\\
$\mu _{\alpha}$ [mas/yr] & $44.2 \pm 3.5^{(c)}$\\ 
$\mu _{\beta}$ [mas/yr] & $6.8 \pm 3.5^{(c)}$\\
$U_{LSR}$ $[\rm km \, s^{-1}]$ & $-154.4 \pm 50.4^{(d)}$\\
$V_{LSR}$ $[\rm km \, s^{-1}]$ & $-63.0 \pm 17.9^{(d)}$\\
$W_{LSR}$ $[\rm km \, s^{-1}]$ & $-48.5 \pm 19.0^{(d)}$\\

\hline
\hline
\end{tabular}
\tablefoot{$^{(a)}$ \cite{2012yCat.1322....0Z}; $^{(b)}$ \cite{2003yCat.2246....0C}; $^{(c)}$ \cite{2000A&A...355L..27H}; $^{(d)}$ This work (see text)}.
\end{table}


\section{Stellar properties}
\label{Stellar properties}

TYC 4282-605-1 is a K giant star with a visible magnitude of V=$10.581 \pm 0.01$ mag (stellar parameters summarised in Table \ref{stellar_parameters_1}) misclassified as an M star in \cite{2011AJ....142..138L} and for this reason it was included in the original M stars sample in the GAPS programme. The first collected spectra showed soon its earlier type and low gravity. Notwithstanding that the star was misclassified, we decided to continue monitoring this target because of the large variations of RV observed in the first few observations. 

Our HARPS-N data (see Sect. \ref{The HARPS-N observations}) were used to determine the stellar parameters through the higher signal-to-noise ratio (S/N) co-added spectrum with a value of $\sim$ 400 at 5500 \AA. Our analysis yielded a temperature of $T_{\rm eff} = 4300 \pm 50$ K, a surface gravity $\log g=2.0 \pm 0.2$ $\rm dex$, a microturbulent velocity $\xi=1.19 \pm 0.2$ $\rm km\, s^{-1}$ and an iron abundance of $\rm [Fe/H]= -0.07 \pm 0.16$ dex with a method based on measurements of equivalent widths (EWs) that relies on the excitation and ionisation equilibria of Fe lines, explained accurately in Sect. \ref{Chemical analysis}. 

To measure the projected rotational velocity ($v \sin i$), we performed a spectral synthesis using the {\it synth} driver within MOOG code \citep[version 2014,][]{1973ApJ...184..839S} and fixing the stellar parameters ($T_{\rm eff}$, $\log g$, $\xi$) at the values derived through the EWs method. We followed the prescriptions given by \cite{2011A&A...526A.103D}, and fixed the macroturbulence velocity to the value derived from the relation by \cite{2005ApJS..159..141V}. Further details on the procedure of spectral synthesis are given in other works within the GAPS project \citep[see, e.g.][]{2013A&A...554A..28C,2015A&A...575A.111D,2014A&A...564L..13E}. The resulting $v \sin i$ is reported in Table \ref{stellar_properties}.

\begin{figure}[h]
\centering
\includegraphics[width=\hsize]{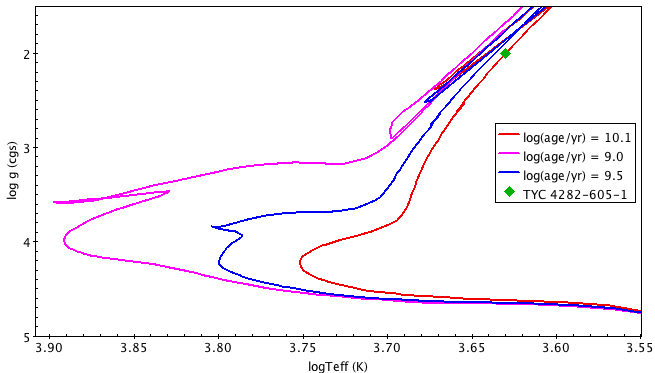}
\caption { \scriptsize Isochrones by \cite{2012MNRAS.427..127B} place the target (green diamond) as an old giant star with log (age/yr) = 10.1 (red line). The rest of the estimated parameters (mass, radius, luminosity) are listed in Table. \ref{stellar_properties} }
\label{isochrone_KG7}
\end{figure}

For the estimation of the stellar mass, radius, luminosity and age (listed in Table \ref{stellar_properties}) we used the isochrones by \cite{2012MNRAS.427..127B} shown in Fig. \ref{isochrone_KG7}. From the spectroscopic parameters ($T_{\rm eff}$ and log $g$) extracted in this work we identified the evolutionary track followed by TYC 4282-605-1, classifying this object as a 10.1 Gyr star, being therefore an old giant star with mass $\rm 0.97 \pm 0.03 M_{\odot}$.

Thanks to the empirical calibration of $T_{\rm eff}$ versus colour and $\rm [Fe/H]$ of giant stars proposed by \cite{1999A&AS..140..261A} we obtained the bolometric correction ($BC(\rm V)=-0.636 \pm 0.003$, as a function of $T_{\rm eff}$ and $\rm [Fe/H]$) and the colour excess ($E_{\rm B-V}= 0.24 \pm 0.01$) that allowed us to estimate the distance of the star according to the following equations:

\begin{equation}
M_{bol}=M_{bol_{\odot}}-2.5log\frac{L}{L_{\odot}} \,\,\, ; \,\,\,\, M_{bol_{\odot}}=4.75 
\end{equation}

\begin{equation}
M_{bol}=M_{V}+BC(V)
\end{equation}

\begin{equation}
M_{V}=m_{V}+5-5logd(pc)-A_{V}
\end{equation}

where the total extinction, $A_{\rm V}$, is defined as $A_{\rm V}= 3.1 \times E_{B-V}$, $M_{\rm V}$ is the absolute magnitude in V band, $m_{\rm V}=10.581 \pm 0.01$ is the apparent V magnitude provided by the UCAC4 catalogue and $d$ is the distance of the star measured in parsec. The computation yields the star at a distance of $700 \pm 143$ pc from the sun.

Since the star has been observed by Gaia and included in the first data release \citep{2016A&A...595A...1G}, we have computed the stellar mass using its Gaia parallaxes of $\pi= 2.74 \pm 0.56$ mas using the relation $g=G \frac{M}{R^2} $ where $g$ is measured spectroscopically and $R$ is computed from $L=\sigma T_{eff}^4 4 \pi R^2$. The resulting mass is $0.26 \pm 0.15$ $M_{\odot}$. Such low mass in unexpected for an evolved giant unless the star has lost substantial fraction of it mass. In this case we should find chemical anomalies that are not evident from our analysis (see Sect. \ref{Chemical analysis}). Since the discrepancy of our parallax with the Gaia one is at only the 2$\sigma$ level, before applying the Lutz-Kelker correction, we assume a stellar distance of $d=700 \pm 143$ pc as determined by isochrones.

\begin{table}[h]
\centering
\caption{Stellar parameters of TYC 4282-605-1 derived in this work.}
\label{stellar_properties}
\begin{tabular}{l c c c c c c c}
\hline
\hline
\noalign{\smallskip}
Parameters  & Value  \\
\noalign{\smallskip}	
\hline	
\noalign{\smallskip}

\textit{Derived from HARPS-N spectra} & &\\
$T_{\rm eff}$ (K)  & $\rm 4300 \pm 50$\\
log $g$ ($\rm dex$) & $\rm 2.0 \pm 0.2$\\
$\xi$ ($\rm km\,s^{-1}$) & $\rm 1.19 \pm 0.2$\\ 
$\rm [Fe/H]$ (dex) & $\rm -0.07 \pm 0.16$ \\
$ v \sin i$ ($\rm km\,s^{-1}$) & $\rm 3.0 \pm 0.5$\\
\hline

\textit{Estimated from isochrones $^{(a)}$} & \\
Mass ($\rm M_{\odot}$) & $\rm 0.97 \pm 0.03$ \\
Radius ($\rm R_{\odot}$) & $\rm 16.21 \pm 4.01$\\
log L ($\rm L_{\odot}$) & $\rm 1.91 \pm 0.18$\\
$d$ (pc) & $\rm 700 \pm 143$ \\
Age (Gyr) & $\rm 10.1 \pm 0.05$\\
\hline
\hline
\end{tabular}
\tablefoot{$^{(a)}$ \cite{2012MNRAS.427..127B}}
\end{table}


\subsection{Chemical analysis}
\label{Chemical analysis}

Chemical abundance of individual elements C, Na, Mg, Al, Si, Ca, Ti, Cr, Fe, Ni, Cu, Y, Ba, and La were obtained using the MOOG code together with ATLAS9 atmosphere models \citep{1993KurCD..13.....K}, with no overshooting. Abundances of Na, Mg, Al, Si, Ca, and Ni were obtained using lines of the neutral atoms (X~{\sc i}), while for Ti, Cr, and Fe we have exploited lines of both neutral and single ionised species (X~{\sc ii}). The lines selected in the chemical analysis of these elements were taken from \cite{2015A&A...579A..20M}. 

For the above mentioned species we have carried out EW analysis, using the driver {\it abfind} in MOOG; the ARES code \citep{2007A&A...469..783S} has been used to measure EW values, but we have carefully double checked each single spectral feature using the task $splot$ in IRAF\footnote{IRAF is the Image Reduction and Analysis Facility, a general purpose software system for the reduction and analysis of astronomical data. IRAF is written and supported by National Optical Astronomy}. For C, Cu, Y, Ba and La we have instead performed spectral synthesis calculations using the driver {\it synth}, and including hyperfine structure and isotopic splitting, as needed. The abundance of C ~{\sc i}  has been determined by synthesing the CH band at $\rm 4300\, \AA$ using the line list by Plez (private communication), whereas
for Cu we have used the line at 5782 \AA, including HFS information by \cite{1985A&AS...59..403S}. As for neutron-capture elements we have employed lines at 4883.68 \AA~ and 4900.12 \AA~ for Y~{\sc ii}, the line at 5853.69 \AA~ for Ba{\sc ii}, and La{\sc ii} lines at 4322.51 \AA~ and 6390.48 \AA. We refer the reader to our previous papers for details on oscillator strengths and atomic parameters for these spectral features \citep[e.g.][]{2012MNRAS.423.2789D,2017A&A...598A..19D}.

The first step is the determination of the atmospheric parameters and
iron abundance by means of the spectroscopic analysis, following the
standard procedure. Effective temperature (T$_{\rm eff}$) has been derived
by zeroing the slope between  abundances from
Fe {\sc i} lines and the excitation potential of the spectral features.
Similarly, microturbulence values ($\xi$) has been obtained imposing no spurious trend between abundances from Fe {\sc i} and the reduced EWs (that is EW/$\lambda$). The surface gravity (log$g$) comes from the ionisation balance, that is $\Delta$[A(Fe~{\sc ii})$-$ A(Fe~{\sc i}) ]=0. The solution is reached when all the three conditions are simultaneously
satisfied, better than 1$\sigma$ from the error on slopes for temperature
and microturbulence and better than roughly one-third
the error bar in Fe{\sc i} and Fe{\sc ii} features (i.e. the standard deviation
from the mean). 
We have obtained T$_{\rm eff}$=4300$\pm$50 K,  log$g$=2.00$\pm$0.2 dex, $\xi$=1.19$\pm$0.20 km~s$^{-1}$. By adopting these atmospheric stellar parameters we have 
inferred an iron abundance of A(Fe~{\sc i})=7.43$\pm$0.01 (rms=0.134, 207 lines) and A(Fe~{\sc ii})=7.43$\pm$0.03 (rms=0.102, 15 lines).\footnote{The derived abundances are expressed in the usual scale, $\rm A(X)=log(N_{X}/N_{H}) + 12$. }

Our results are reported in Table \ref{abundancias}, which includes abundances for the species under scrutiny in this study along with corresponding uncertainty. 
The solar abundances adopted throughout the manuscript are given in Column 2 and have been employed to derive the [X/Fe] ratios listed in Column 5.\footnote{[X/Fe]=[X/H]$-$[Fe/H], where [X/H]=A(X)-A(X)$_\odot$.}
Two kind of internal (random) errors affect our abundance values, that is errors due to EW measurements (or to the determination of best fit via spectral synthesis) and errors related to atmospheric parameters.
For the first source of errors, assuming Gaussian statistics, the line-to-line scatter errors are computed as $\rm \sigma/\sqrt{n}$, where $\sigma$ is the standard deviation of the derived individual abundances from the $n$ lines. For species for which only one spectral line is available (i.e., Cu and Ba) we have repeated the measurement
several times by changing the continuum displacement
and other criteria and inspected the corresponding variation
in the resulting abundances.
In order to estimate uncertainties due to stellar parameters we have instead proceeded in the standard way, that is by varying one parameter at the time and inspecting
the corresponding change in the resulting abundance.
The total uncertainties for the derived abundances are then calculated by summing in quadrature line-to-line scatter errors and those related to the stellar parameters (Column 5).

Our findings point to a significant enrichment in the abundances of Na~{\sc i}, Mg~{\sc i}, Al~{\sc i} and Si~{\sc i} while Ca~{\sc i}, Ti~{\sc i},~{\sc ii}, Cr~{\sc i},~{\sc ii} and Ni~{\sc i} are consistent with the solar value. The same result holds for heavy elements (Cu~{\sc i}, Y~{\sc ii}, Ba~{\sc ii} and La~{\sc ii}) for which we retrieve a solar-scaled abundance pattern.

\begin{figure}[t]
\centering
\includegraphics[width=\hsize]{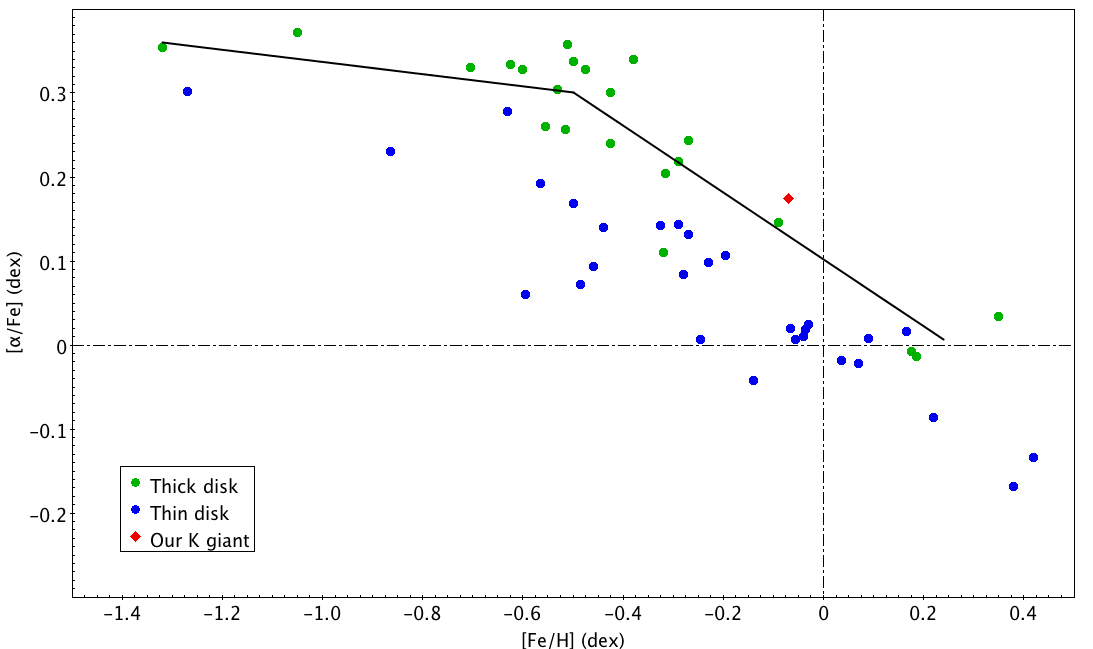}
\caption { \scriptsize Mean $\rm \alpha$-elements abundance ratio ([O, Mg, Si, Ca, Ti/Fe]) as a function of [Fe/H] by \cite{2010A&A...513A..35A}. The symbols are explained in the figure. }
\label{Alves-Brito_thickdisk}
\end{figure}

To investigate the nature of the enhancement of some of the studied elements we decided to compare our results with the work by \cite{2010A&A...513A..35A} for a sample of giant stars in the local disk. The authors show a chemical distinction between the local thin and thick disks stars. Their results suggest that the local thick disk stars have $\rm \alpha$-enhancements relative to solar abundances in K giants higher than those stars located in the local thin disk. As shown in Fig. \ref{Alves-Brito_thickdisk} the $\rm [\alpha/Fe]$ chemical pattern of TYC 4282-605-1 clearly follows the performed linear fit for $\rm [\alpha/Fe]$ vs. $\rm [Fe/H]$ of local thick disk stars. We demonstrated thus that the chemical composition of our target supports that it is an old K giant (age = 10.1 Gyr) belonging to local thick disk.

This result is confirmed by the kinematic velocities. Calculation of the space velocity with respect to the Sun is based on the procedure presented by \cite{1987AJ.....93..864J}, corrected for the effect of differential galactic rotation \citep{1988Sci...240.1680S}, by adopting a solar Galactocentric distance of 8.5 kpc and a circular velocity of 220 $\rm km \, s^{-1}$. The correction of space velocity to the local standard of rest is based on a solar motion, $(U, V, W)_{\odot}=$ (10.0, 5.2, 7.2) $\rm km \, s^{-1}$, as derived from Hipparcos data by \cite{1998MNRAS.298..387D}. TYC 4282-605-1 shows kinematic properties typical of the thick disk population (see. Table \ref{stellar_parameters_1}) following the distributions of the space velocities for the thick disk sample calculated by \cite{2004A&A...421..969B}.

\begin{table*}[h]
\centering
\caption{Derived abundances for TYC 4282-605-1}
\label{abundancias}
\begin{tabular}{l c c c c}
\hline
\hline
\noalign{\smallskip}
Ion & $\rm \langle A(X)\rangle_{\odot}^{(1)}$  & $\rm \langle A(X)\rangle^{(1)}$ & n  &  $\rm \left[ X/Fe \right]^{(2)}$ \\

\noalign{\smallskip}	
\hline	
\noalign{\smallskip}


C~{\sc i}  & $8.43 \pm 0.07 $  &  $8.20\pm0.08 $    & 1  &  $-0.23 \pm 0.15$ \\
Na~{\sc i} & $6.38 \pm 0.01$ & $6.57 \pm 0.04$ &  3    & $ 0.26  \pm 0.16$\\
Mg~{\sc i} & $7.62 \pm 0.02$ & $7.91 \pm 0.04$ &  2    & $ 0.36  \pm 0.13$\\
Al~{\sc i} & $6.48 \pm 0.01$ & $6.82 \pm 0.09$ &  2    & $ 0.41  \pm 0.15$\\
Si~{\sc i} & $7.60  \pm 0.01$ & $7.78 \pm 0.03$ &  13  & $ 0.25  \pm 0.05$\\
Ca~{\sc i} & $6.42 \pm 0.02$ & $6.46 \pm 0.03$ &  12   & $ 0.11  \pm 0.18$\\
Ti~{\sc i} & $5.02 \pm 0.01$ & $4.97 \pm 0.02$ &  15   & $ 0.02  \pm 0.17$\\
Ti~{\sc ii}& $5.04 \pm 0.01$ & $4.91 \pm 0.05$ &  5    & $ -0.06 \pm 0.06$\\
Cr~{\sc i} & $5.68 \pm 0.01$ & $5.63 \pm 0.03$ &  18   & $ 0.02  \pm 0.16$\\
Cr~{\sc ii}& $5.67 \pm 0.01$ & $5.67 \pm 0.06$ &  3    & $ 0.07  \pm 0.06$\\
Ni~{\sc i} & $6.29 \pm 0.01$ & $6.28 \pm 0.02$ &  37   & $ 0.06  \pm 0.05$\\
Cu~{\sc i}   & $4.25 \pm 0.05$  &  $4.19 \pm 0.07$   &  1  &  $0.01  \pm 0.15$   \\
Y~{\sc ii}     & $2.19 \pm 0.05$ &  $2.06 \pm 0.04$    &  2  &  $-0.06 \pm 0.20$   \\
Ba~{\sc ii}   & $2.13 \pm 0.05$ &  $2.08 \pm 0.08$    &  1  &  $0.02  \pm 0.25$   \\
La~{\sc ii}   & $1.05 \pm 0.04$ &  $0.97 \pm 0.01 $   &  2  &  $-0.01 \pm 0.20$    \\

\noalign{\smallskip}

\hline
\hline
\noalign{\smallskip}

 & $\rm \langle A(X)\rangle_{\odot}^{(1)}$  & $\rm \langle A(X)\rangle^{(1)}$ & n & $\rm \left[ Fe/H \right]^{(2)}$ \\
\noalign{\smallskip}	
\hline	
\noalign{\smallskip}


  Fe  	      & $7.50 \pm 0.01$ & $7.43 \pm 0.02$ &  222  & $-0.07 \pm 0.19$\\

\hline
\hline
\end{tabular}
\tablefoot{$^{(1)}$ line-to-line scatter errors as $\rm \sigma \sqrt{n}$, $^{(2)}$ final uncertainties due to the propagation of errors in stellar parameters plus the line-to-line scatter errors}
\end{table*}

\section{HARPS-N observations and radial velocity analysis}
\label{The HARPS-N observations}

We collected 48 spectra of TYC 4282-605-1 with HARPS-N at TNG in the four seasons from August 2012 to November 2015. Spectra were obtained with an integration time of 900 s and an average S/N of 70 at 5500 \AA. The main characteristics of HARPS-N, very similar to HARPS at ESO, are the very high stability, the high spectral resolution of R=115,000 and the large wavelength coverage from 3800 to 6900 \AA.

\begin{figure}[h]
\centering
\includegraphics[width=\hsize]{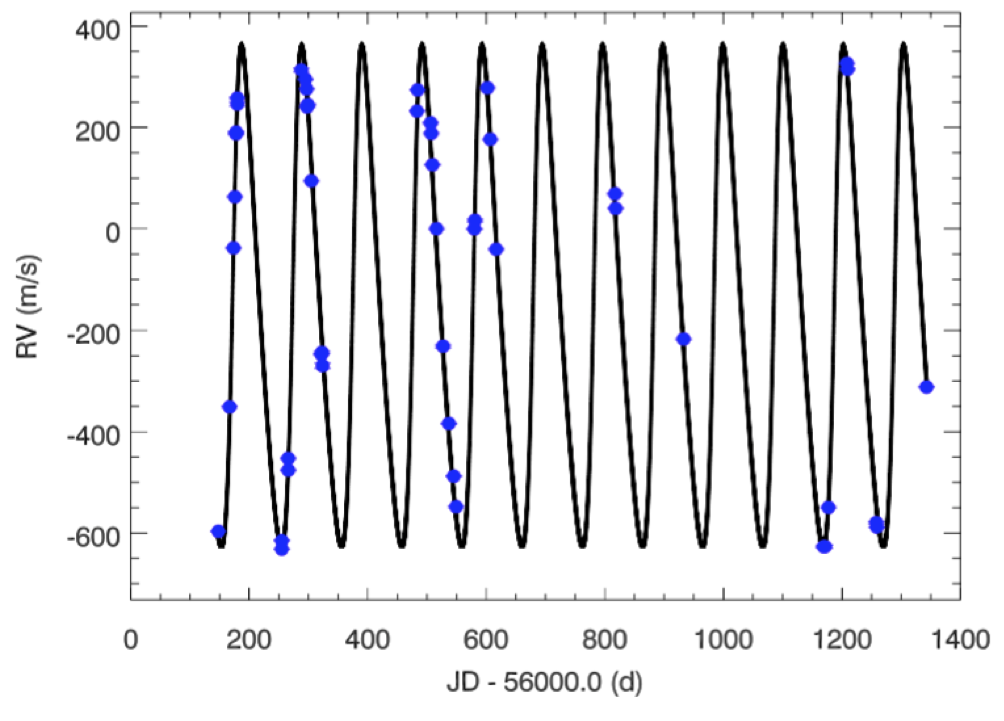}
\caption { \scriptsize Radial velocity measurements from the four observing seasons with HARPS-N (blue dots). Removal of median of RV has been applied to the data set before plotting. The curve represents the orbital solution (see Table \ref{parameters}). }
\label{RV_time_HARPS}
\end{figure}

The reduction of the spectra and the RV measurements were obtained using the latest version (Nov. 2013) of the HARPS-N instrument Data Reduction Software (DRS) pipeline with a K5 binary mask. The measurement of the RVs is based on the weighted cross-correlation function (CCF) method \citep{1996A&AS..119..373B,2002A&A...388..632P}. In Table \ref{Harps_rv_measurments}, we list the RVs and their corresponding errors. We obtained the first three spectra with the simultaneous Th-Ar calibration lamp, while the rest of the spectra, obtained with a new charge-coupled device (CCD) after the failure of the previous one, were collected with the sky background in the second fiber. We have verified that our results do not change considering or excluding the first three data points, and that a systematic correction offset is not needed. The RV measurements for the four seasons are shown in Fig. \ref{RV_time_HARPS} after removal of median RV with a standard deviation of 337.5 $\rm ms^{-1}$ much more larger than the average error (1.8 $\rm ms^{-1}$), pointing to an intrinsic variability. 

To detect a possible periodic signal in our data we used the Generalized Lomb-Scargle periodogram \citep[GLS,][]{2009A&A...496..577Z}, which is the most common method in case of unevenly spaced times and it is characterised by a simple statical behaviour. The GLS periodogram allows us to identify significant periods in the data which can be used as starting estimates for the algorithm if their amount of power is higher than a confidence level. 

\begin{figure}[h]
\centering
\includegraphics[width=\hsize]{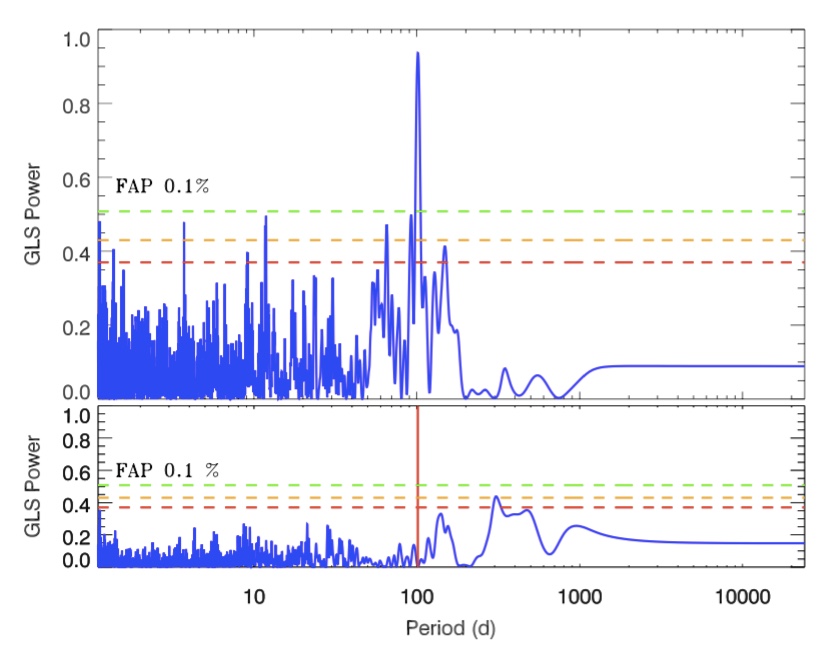}
\caption { \scriptsize \textbf{Top panel:} GLS periodogram of RVs measured with HARPS-N with a significant peak at 101.54 days, the horizontal dashed lines represent the false alarm probabilities (FAP) of 0.1$\%$ (green line), 1$\%$ (orange) and 10$\%$ (red), respectively. \textbf{Botton panel:} Periodogram of the residuals data after subtracting the orbital fit. The red solid line indicates the location of the maximum period found with the RV. } 
\label{KG7_period_harps}
\end{figure}

\begin{figure}[h]
\centering
\includegraphics[width=\hsize]{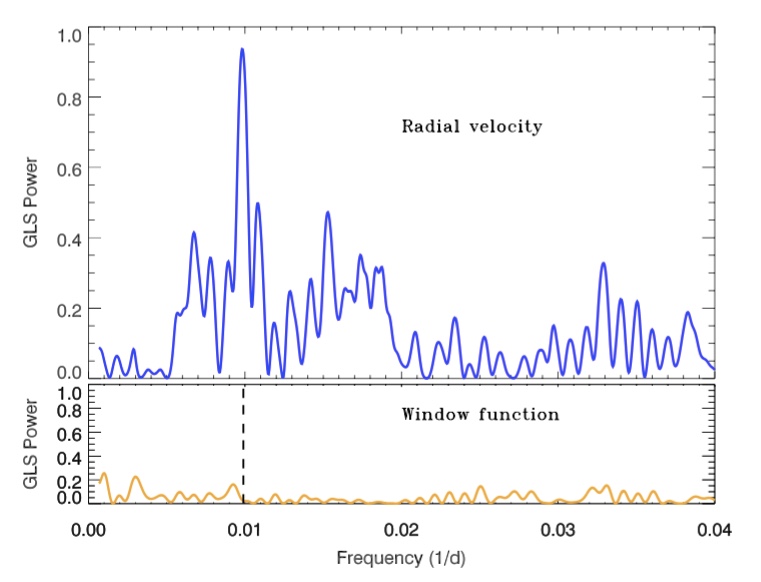}
\caption { \scriptsize GLS periodogram of the TYC 4282-605-1 RVs (upper panel) and the corresponding window function (lower panel). The strong peak found around 101 days corresponds with the value in frequency of 0.01 $\rm days^{-1}$. }
\label{KG7_window_function}
\end{figure}

The GLS periodogram reveals a clear periodicity with a highly significant peak at period $P=101.54$ days in the top panel of Fig. \ref{KG7_period_harps} with a power much higher than the lowest false alarm probability (FAP) fixed at 0.1$\%$ (green line) and estimated through a bootstrap method with 10,000 iterations. To verify the presence of possible aliasing phenomenon, we plot the window function in Fig. \ref{KG7_window_function} showing that the strong peak around 101 days, corresponding to 0.01 $\rm days^{-1}$ in the domain of frequencies, is not related with a periodicity in the sampling. 

Assuming the companion hypothesis we used the RVLIN code \footnote{The code is available at http://exoplanets.org/code/} \citep[see][]{2009ApJS..182..205W} to obtain the best Keplerian fit and derive the orbital parameters listed in Table \ref{parameters}. The orbital solution and RV data are presented in Fig. \ref{RV_time_HARPS} as a function of the time. 

\begin{figure}[h]
\centering
\includegraphics[width=\hsize]{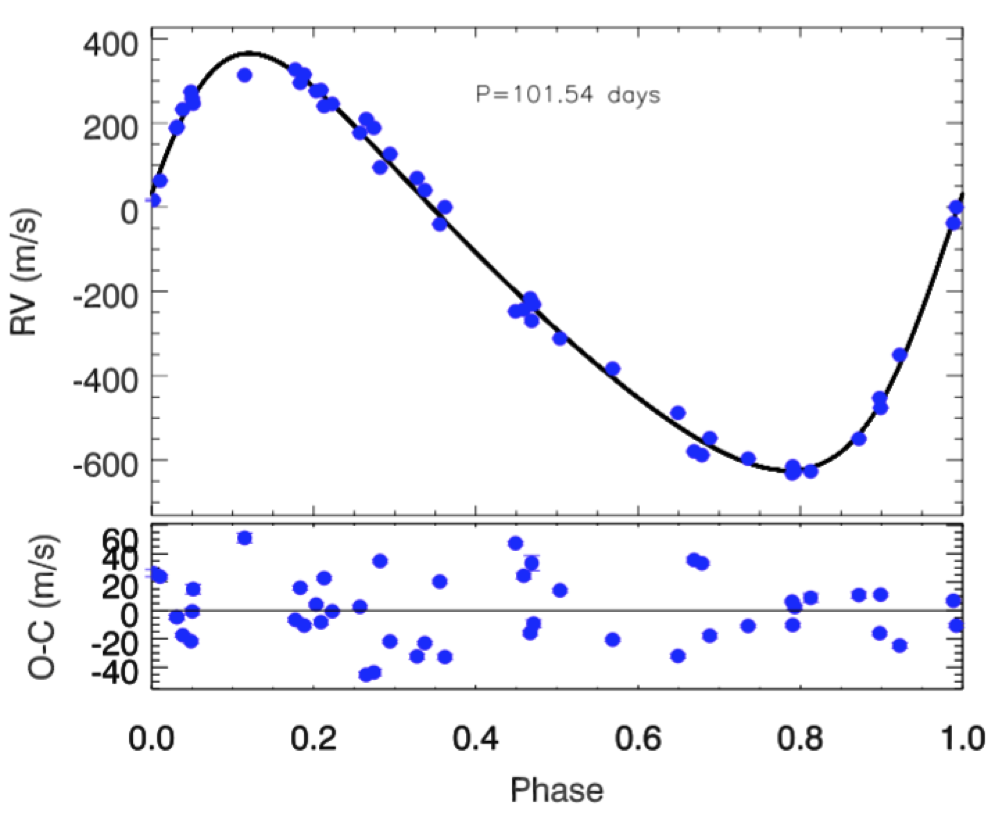}
\caption { \scriptsize \textbf{Top panel:}Radial velocity data of the four observing seasons of HARPS-N phased to the correspond orbital period. The solid line is the orbital solution. \textbf{Bottom panel:} Residual RV variations after subtracting the orbital solution. The residuals show standard deviation value of 23.02 $\rm ms^{-1}$.}
\label{KG7_phase_harps}
\end{figure}

In order to obtain the residuals data we use the orbital fit to remove the period and its harmonics simultaneously. The periodogram of the residuals (bottom panel of Fig. \ref{KG7_period_harps}) does not show other significant peaks, but a marginal long term signal between 300-500 days is present. The RV and residuals measurements (rms = 23.02 $\rm ms^{-1}$) phase-folded are shown in Fig. \ref{KG7_phase_harps}. Our solution resulted in a semi-amplitude, $K_{\rm opt}=495.2\pm5.1$ $\rm ms^{-1}$, eccentricity, $e=0.28\pm0.01$, semi-major axis, $a=0.422 \pm 0.009$ $\rm AU$ and a minimum mass for the companion, $M_P\sin i = 10.78\pm 0.12 \; M_{\rm J}$, assuming the stellar mass of $\rm 0.97 \pm 0.03$ $\rm M_{\odot}$ previously derived. We estimated the errors of the derived orbital parameters through a bootstrap (10,000 re-sampling) analysis.

\begin{table*}[h]
\centering
\caption{Best fit orbital parameters}
\label{parameters}
\begin{tabular}{l l c }
\hline
\hline
\noalign{\smallskip}
Parameter & Units & Value \\
\noalign{\smallskip}	
\hline	
\noalign{\smallskip}

$P$\dotfill & Period (days)\dotfill & $101.54 \pm 0.05$\\
$T_P$ \dotfill & Periastron time (JD-2,400,000)\dotfill & $56479.28 \pm 0.57$\\
$e$\dotfill & Eccentricity\dotfill & $0.28 \pm 0.01$\\
$\omega_*$\dotfill & Argument of periastron (deg)\dotfill& $288.94 \pm 2.38$\\
$K$\dotfill & RV semi-amplitude (m/s)\dotfill & $495.2 \pm 5.1$\\

$\gamma$\dotfill & Systemic velocity (m/s)\dotfill & $-12,361.54 \pm 3.58$\\
$M_P\sin i$\dotfill & Minimum mass ($\rm M_{J}$)\dotfill & $10.78 \pm 0.12$\\
$a$\dotfill & Semi-major axis (AU)\dotfill & $0.422 \pm 0.009$\\

\hline
\end{tabular}
\end{table*}


\subsection{Bisector analysis}

\begin{figure}[h]
\centering
\includegraphics[width=\hsize]{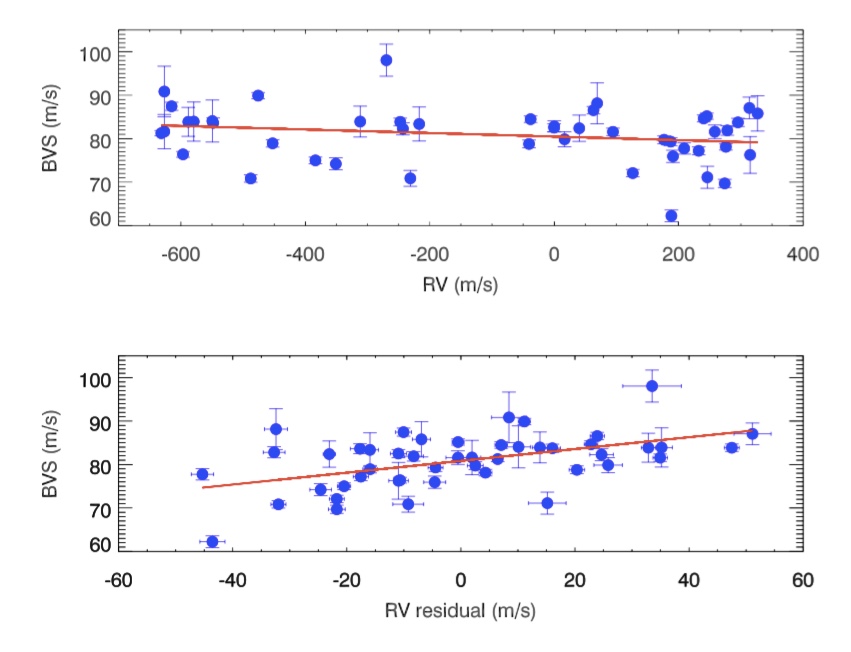}
\caption { \scriptsize Correlation between bisector velocity span (BVS) with RV measurements (top panel) and with residuals of RV (bottom panel). }
\label{KG7_BVS_RV_and_res}
\end{figure}

Stellar activity and pulsations can cause deformations in the line profile of the spectral lines producing a shift of the centroid of the line. This effect can be quantified by several asymmetry indicators, including the bisector. In presence of surface inhomogeneities or surface deformations we expect some (anti) correlation between RV and the CCF bisector velocity span (BVS). The top panel of Fig. \ref{KG7_BVS_RV_and_res} shows the BVS as a function of the RV measurements. No significant correlation was found with Spearman correlation coefficient of $\rho = -0.17$. The bottom panel of the same figure shows the correlation between BVS and RV residuals having a value of $\rho = 0.47$. In Fig. \ref{KG7_Sindex_BVS_res_GLS} the GLS periodogram of the BVS is shown and no significant peak is found at the 101 days period, indicating that the modulation found in the RV is probably not connected to stellar activity or pulsations. That instead may be responsible for the long term (> 300 days) signal and for the correlation between BVS and RV residuals likely related to the stellar rotation. The lack of a correlation does not imply necessarily the presence of a companion as the origin of the RV periodicity but it is a necessary condition.

\subsection{Chromospheric emission, S-index}

Stellar activity can produce RV variations and create a signal that could be misinterpreted as an exoplanet. Some apparent RV offsets are related to the magnetic activity of the star and surface features such as spots. Also, the convection patterns of the stellar surface can suffer local or global changes due to the magnetic activity. When a star has a large convection layer hotter and bluer material can emerge from its convection cells (e.g. stellar granulation) and it causes apparent blueshifts to the integrated stellar spectrum \citep{2012ApJS..200...15A}. For these reasons, one could expect apparent RV jitter whenever the magnetic field of the star experiences changes. One way to disentangle the effects of activity is to measure activity indicators such as the Mount Wilson S-index \citep{1984ApJ...279..763N}. This index measures the relative flux of the Ca II H and K lines in emission ($\rm \lambda_{K} = 3933.664 \, \AA \, and \, \lambda_{H} = 3968.470 \, \AA$) compared with a local continuum \citep[e.g.][]{2011arXiv1107.5325L}. These lines in emission are formed in the hot plasma of the chromospheres of stars, and their intensity varies with the strength of the stellar magnetic field. 

The width of the chromospheric lines in giant stars is larger than in main-sequence stars because of the Wilson-Bappu effect \citep[e.g.][]{1980ARA&A..18..439L,2013AJ....146...73P}. Nevertheless, their width can still be sensitive to the activity level of the  star, that is, to the level of non-thermal heating of its chromosphere, as found by \cite{1997A&A...326..165E}. Therefore, we shall use the chromospheric S-index to look for a possible correlation with the RV variation that can indicate activity as the physical cause of the observed RV modulation, although the sensitivity of the method is lower than in the case of main-sequence stars. We checked that the band passes defined by \cite{2011arXiv1107.5325L} for main-sequence star are adequate for our giant and we used HARPS-TERRA (Template-Enhanced Radial velocity Re-analysis Application) software \citep{2012ApJS..200...15A} that incorporates the automatic measurements of the S-index and its corresponding errors using the necessary information provided by the HARPS DRS. 

\begin{figure}[h]
\centering
\includegraphics[width=\hsize]{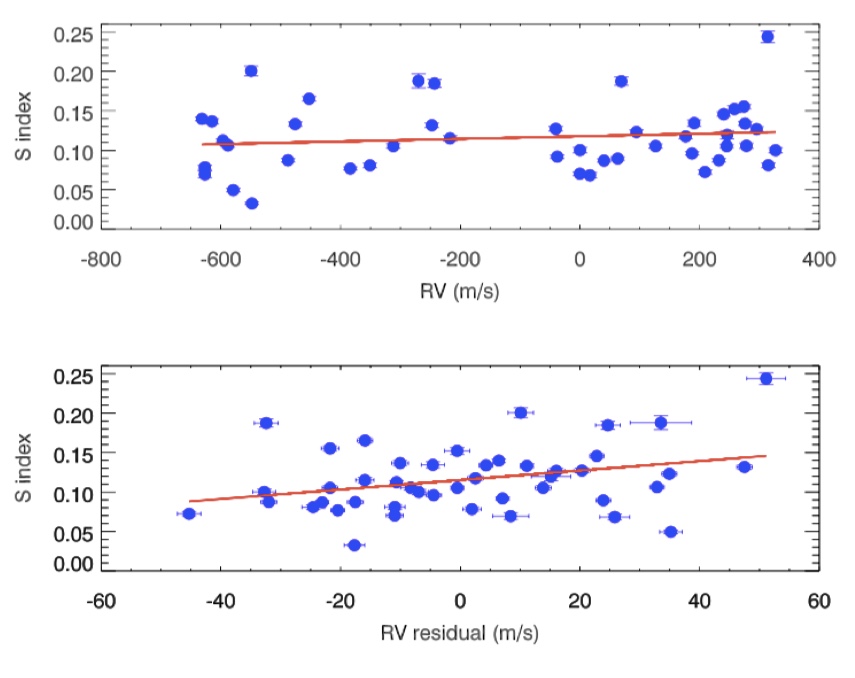}
\caption { \scriptsize Correlation between RV (top panel) and RV residual (bottom panel) with S-index.}
\label{Sindex_rv_and_res_2}
\end{figure}

\begin{figure}[h]
\centering
\includegraphics[width=\hsize]{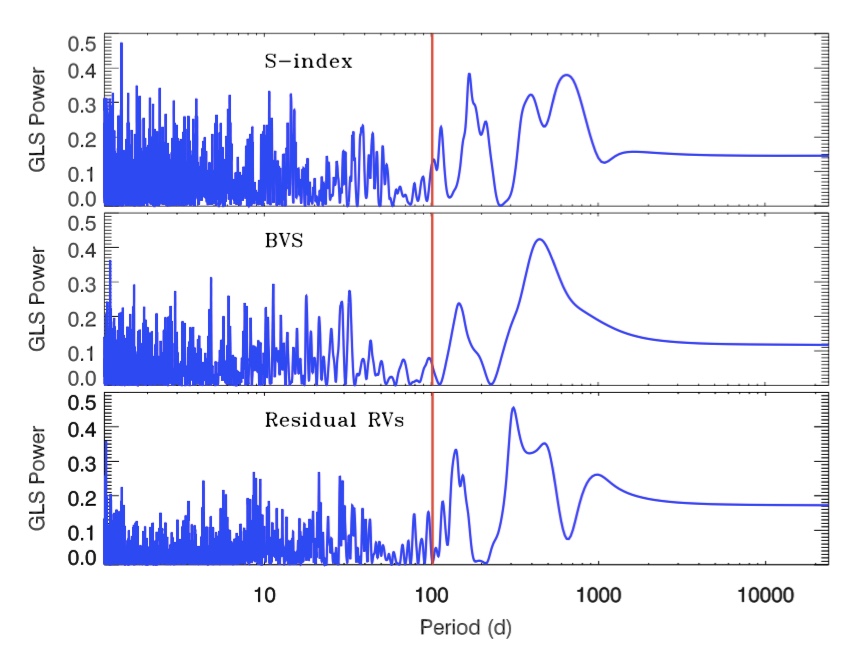}
\caption { \scriptsize GLS Periodograms of S index, BVS and residual RV. There is no significant period at 101 days (red line).}
\label{KG7_Sindex_BVS_res_GLS}
\end{figure}

Another argument in favour of the companion interpretation at 101 days is shown in Fig. \ref{Sindex_rv_and_res_2} through the low Spearman correlation coefficient between the S index with RV ($\rho = 0.15$, in the top panel) and with the RV residuals ($\rho = 0.29$, middle panel), respectively. Figure \ref{KG7_Sindex_BVS_res_GLS} shows the GLS periodograms for S index, BVS and residual RVs in order to check that there is not activity effects correlated at 101 days. As a result, the activity has been excluded as being responsible of the significant signal found in optical band.

\subsection{The case of a pulsating star}
\label{The case of a pulsating star}

The physical parameters listed in Table \ref{stellar_properties} and the RV curve shown in Fig. \ref{KG7_phase_harps} suggest the possibility that TYC~4282-605-1 is a pulsating star.
Therefore, we must discuss it on the basis of what
described in Sect. \ref{Introduction}. 

We note that the RV curve has a reverse
shape of that expected from a radial pulsation mode \citep[see
Fig.~2 in][for a close comparison with the
RV curve of $\delta$~Cep]{2017A&A...597A..73N}.

A rising branch steeper than the descending one in the
RV curve implies a descending branch steeper than the
rising one in the photometric light curve, but such a shape
is not very common in pulsating stars. These light curves seem
to be confined to few unusual cases, like V1719~Cyg stars.  
However, these variables show only slight
asymmetric RV curves \citep[see Fig.~3 in][]{1988A&A...199..191P}
and therefore the RV curve of TYC~4282-605-1 would be unique.

The line profiles have also to change during pulsation, to reflect the outward and inward motion of the atmosphere  \citep[see Fig.~1 in][]{2017A&A...597A..73N}. In particular, variations of the full-width half-maximum (FWHM) and of the bisector span values are expected over the pulsation period.
No periodicity has been observed in the bisector spans of 
TYC~4282-605-1 (Fig. \ref{KG7_Sindex_BVS_res_GLS}). We also analyzed the FWHM values and we did not find any variability as well. 

Therefore,  the detailed analysis of the HARPS-N spectra does not 
support pulsation as the cause of the RV curve shown in Fig. \ref{KG7_phase_harps} upper panel. On the other hand, the scatter observed in the residuals (Fig. \ref{KG7_phase_harps}, lower panel) could be due to solar-like oscillations and/or granulation effects \citep[][]{2011A&A...529L...8K}. For comparative purposes, solar-like oscillations are visible with an amplitude up to 60~m\,s$^{-1}$ and apparent periodicity around 1~d in HD~170053, a red giant very similar to TYC~4282-605-1 \citep{2015ASSP...39..101P}. Unfortunately, the time sampling of the HARPS-N spectra of TYC 4282-605-1 is not suitable for detecting amplitude and time scale of such short-period oscillations.

\section{Photometric analysis}
\label{Photometric analysis}

\begin{figure}
\centering
\includegraphics[width=\hsize]{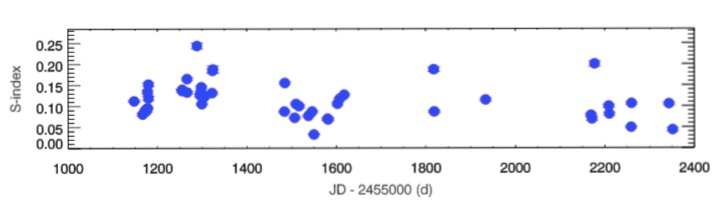}
\includegraphics[width=\hsize]{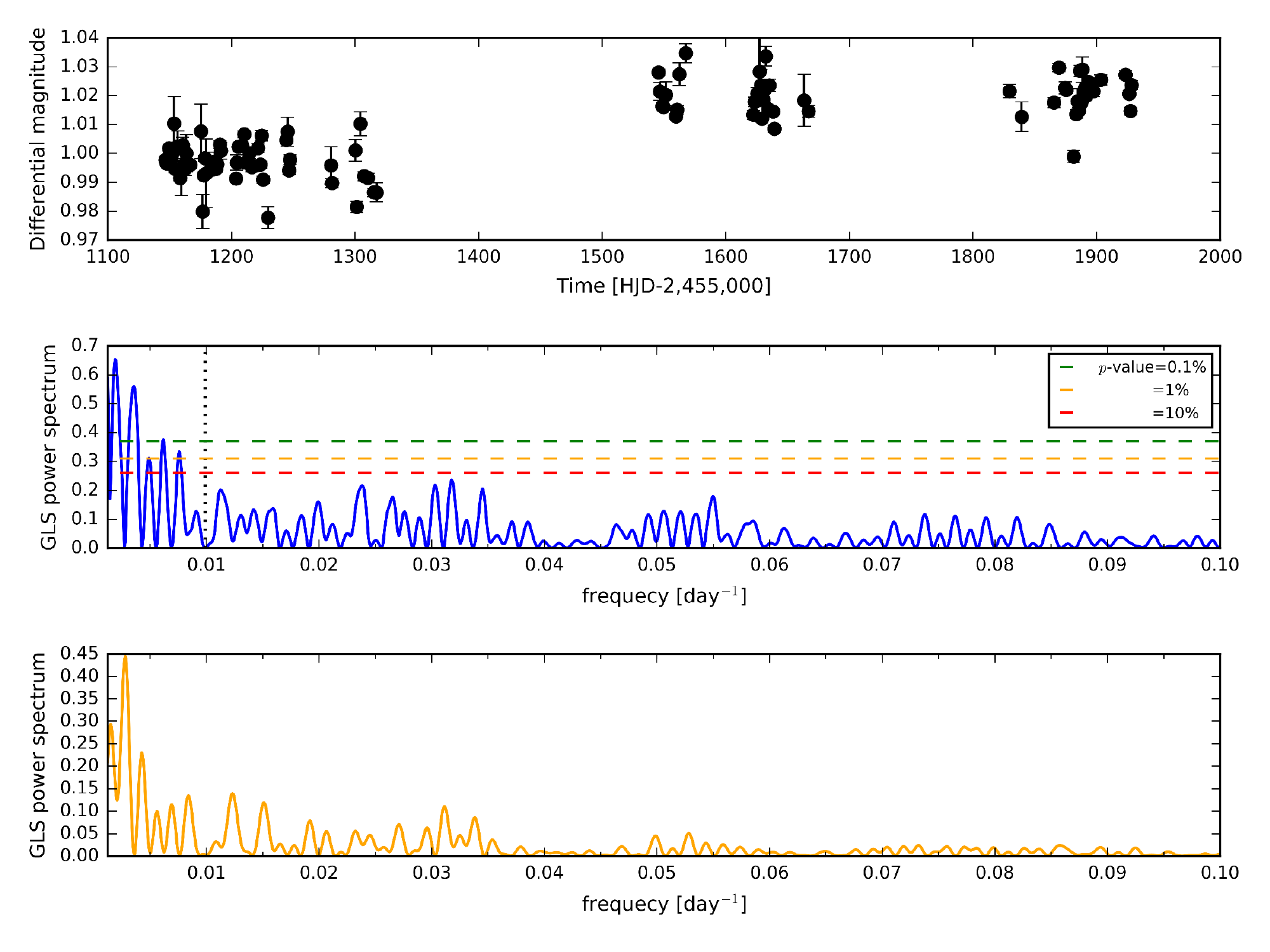}
\caption{\scriptsize \label{fig:apache_full_dataset}\textbf{Upper panel}:Time serie of the S-index. \textbf{Middle-Upper panel}: Differential light curve of TYC 4282-605-1 collected by the APACHE survey. \textbf{Middle-Bottom panel}: Generalized Lomb-Scargle (GLS) periodogram of the differential light curve. The dotted vertical line marks the orbital period of the TYC 4282-605-1 companion. \textbf{Bottom panel}: Spectral window function of the APACHE observations.} 
\end{figure}

Any detected photometric variability on the same time scale of the RV variations makes the companion hypothesis questionable \citep{2002AN....323..392H}. We checked for evidence of a $P \sim$100 days periodicity in the light curve obtained by the APACHE survey \citep[A PAthway to the Characterization of Habitable Earths,][]{2013EPJWC..4703006S}. More than 6350 photometric points in standard Cousin \textit{I}-band have been collected over three observing seasons between 2012 and 2014, with the time span covering a total of 782 days. The differential light curve was derived by using UCAC4 762-068173, 761-069159, 761-069142 and 762-068154 as comparison stars. For the present analysis we have considered the nightly mean values of the total light curve assuming the standard deviation of the mean as the uncertainty for each binned value. We discarded four nights over a total of 109 because of the photometry scarce quality resulting in errors larger than 0.02 mag. We calculate the GLS periodogram in the frequency range 0.00125-0.1 d$^{-1}$ (10-800 days) to search for possible periodic modulation, limiting our investigation to periods lower than the observational time span (Fig. \ref{fig:apache_full_dataset}). The highest peak occurs at $P \sim$ 530 days (0.00189 d$^{-1}$), with \textit{p}-value=0.1$\%$ estimated through a bootstrap (with re-sampling) analysis. Despite the statistical significance, we cannot conclude anything about the physical nature of this signal, but only ascertain the existence of a possible long-term variability, due to the poor sampling of our dataset and the fact that it does not cover the minimum two cycles necessary for verifying the actual periodic modulation. 

However we note that the signal could coincide with the long-term signal found in the activity indicators (bisector, S-index) and present in the residuals of the radial velocity (See Fig. \ref{KG7_Sindex_BVS_res_GLS}) likely linked to the rotation of the star. We note that the light curve (see Fig. \ref{fig:apache_full_dataset}, second panel) presents an offset between the first season and the rest of the observations. The increase occurring between the first and second season seems to correspond to a decrease of S-index (see upper panel of Fig. \ref{fig:apache_full_dataset}). This is suggestive of an activity long-term effect, with higher activity during the first season (low photometry and high S-index) and low activity level during the second and third seasons.

What is more relevant here is that the periodogram, that considers the three observing seasons, does not show significant signals at or close to $P \sim 101$ days. The same result is obtained when the observing seasons have been analysed separately and this also occurs for the residuals (not showed). This lack of photometric signal at 101 days is another point against radial pulsation as origin of the RV variability (see Sect. \ref{The case of a pulsating star}).

\section{GIANO NIR observations}
\label{The GIANO near-IR observations}

\subsection{Observations and data reduction}

We have collected seven NIR spectra from May 2015 to August 2015 (time coverage 100 days) with GIANO at the TNG, with the epochs of observation previously selected to optimise the covering phase found with HARPS-N optical data.

GIANO is the NIR high resolution spectrograph of the TNG that was
originally designed for direct feeding of light at the Nasmyth-B focus \citep{2006SPIE.6269E..19O}. The instrument provides cross dispersed echelle spectroscopy at a resolution of 50,000 over the 0.95-2.45 micron spectral range in a single exposure (Y, J, H and K bands). In July 2012 the spectrometer
was provisionally positioned in Nasmyth-A focus fed via fibers \citep{2014SPIE.9147E..1EO}. The spectrometer was offered to the community in 2015 (TNG call AOT31). In the meantime, a more appropriate interface for direct light-feeding was designed and built, and the spectrograph in the new configuration is now called GIANO-B. This new interface, that also includes a dichroic for simultaneous observations of GIANO-B and HARPS-N \citep[GIARPS,][]{2016SPIE.9908E..1AC}, was succesfully implemented and commisioned between August 2016 and March 2017.

Our observations were collected with the previous configuration of GIANO. It was fiber-fed with two fibers of 1 arcsec angular diameter at a fixed angular distance of 3 arcsec on sky. An image slicer subdivides the light beam from each fiber so the echellogram is composed by four traces instead of two. A typical 2D output of GIANO can be seen and is explained in detail in the Appendix (Fig \ref{2D_output_GIANO}). The GIANO spectrograph is equipped with a 2048 x 2048 HAWAII-2 detector, allowing us to image almost the whole spectral range over 49 orders, with only small missing regions at the longest wavelengths.

Since GIANO is not equipped yet with a gas absorption cell for high precision radial velocities measurements, we obtained the RVs by using the telluric lines as a reference. However, our method can be easily adapted with absorption cells that are now planned in an upgrade of this instrument. 

To extract and wavelength-calibrate the GIANO spectra, we have used the ECHELLE package in IRAF and some new, ad hoc scripts that have been grouped in a package named GIANO TOOLS \footnote{Available on \url{http://www.tng.iac.es/instruments/giano/}}. The GIANO data reduction manual \footnote{\url{http://www.bo.astro.it/giano/documents/handbook_giano_v1.2.0.pdf}} with a detailed description is available on the TNG website and in the Appendix we reported the main steps followed in the reduction of the spectra obtained with GIANO and used in this work.

Observations of science targets are performed by nodding-on-fiber technique, i.e. target and sky are taken in pairs and alternatively acquired on fiber A and B, respectively, for an optimal subtraction of the detector noise and background. From each pair of exposure, an (A-B) 2D-spectrum is computed and then is extracted and summed to get a final 1D wavelength-calibrated spectrum with the best possible signal-to-noise ratio. The spectra in this work were obtained with 600s integration time for each nodding.

2D-spectra of halogen lamps are used to map the geometry of the four spectra in each order, for optimal extraction purposes. The four spectra in each order are independently extracted and wavelength-calibrated, to minimise feature smearing in wavelength due to the small distortion of the slit image along the spatial direction. The instrument is stable enough that flat-fields taken during the daytime are perfectly suited for this purpose.

Each extracted spectrum is wavelength-calibrated by using the U-Ne lamp reference spectra taken at the end of the night. We use a set of approximately 30 bright lines (mostly Ne lines) distributed over a few orders to obtain a first fit, then the optimal wavelength solution is computed by using $\sim$ 300 U-Ne lines distributed over all orders.


\subsection{Radial velocity computations}
\label{Obtaining the radial velocities in near-IR}

In the computation of the RVs in the NIR an ensemble of interface definition language (IDL) procedures was used. It was created to measure RVs with the Cross Correlation Function (CCF) method \citep{1979VA.....23..279B}. A brief summary of the procedure is explained in the following and more details can be found in \cite{2016ExA....41..351C}.

\begin{itemize}

\item Pre-reduction: spectrum normalisation

The first step is to have a uniform wavelength scale of the spectra and it is necessary re-sample the original ones in order to have a constant step in RV and reduce the RV errors. The re-sampling should be made with the same steps for stellar and telluric lines. In order to re-sample the individual order of the input spectrum we make use of a third degree cubic spline interpolation and a continuum normalisation.\\

\item Subtraction of telluric contribution

The goal is to obtain the stellar spectrum cleaned from telluric lines. First  a median spectrum of the Earth atmosphere was created in order to subtract it from the normalised stellar spectra. In the same way, the telluric spectrum cleaned from stellar contribution is obtained. Therefore, for each science observation we obtained two cleaned spectra: the stellar spectrum without telluric lines, and the telluric spectrum without the stellar contribution, both of them are used in order to derive the stellar and telluric RV.\\

\item Stellar mask

We use the CCF method, correlating the spectrum with a mask. It was necessary to prepare two masks, for the stellar and telluric spectra, respectively. It is important that the same mask is used for all the stellar spectra in order to obtain the variations of RVs rather than in their absolute values. From the re-sampled, normalised and cleaned spectra (star and telluric), we built a list of stellar lines and masks.\\

\item Telluric mask

The line list of the telluric target, just like stellar mask, is obtained by considering the median telluric spectrum. Using the line list obtained with the normalised spectra of the telluric standard it is possible to build the telluric mask. We chose only the telluric lines with a similar intensity to the stellar lines.\\

\item Subtraction of stellar contribution

The stellar and telluric spectra can contaminate each other, therefore we subtract the stellar template in order to obtain telluric spectrum without stellar contribution.\\

\item High precision RVs

The subtraction of the telluric RV from the star RV gives the final RV for each spectrum, $\rm RV = RV_{star}- RV_{tell}$ and the internal errors are obtained taking into account the weight of each order and assuming that S/N is given only by statistics of photons. The total error for one specific order is the combination of the telluric and stellar contribution. 

\end{itemize}

The RV errors computed as above vary from epoch to epoch with a mean value about 18 $\rm ms^{-1}$. The final RV and the corresponding error for each epoch is listed in Table \ref{Giano_rv_measurments}.


\begin{table}
\centering
\caption{Measurements of radial velocity with GIANO}
\label{Giano_rv_measurments}
\begin{tabular}{ccc}
\hline
\hline
\noalign{\smallskip}
Epoch [BJD]   & RV $\rm [kms^{-1}]$ & $\rm error \, [\rm kms^{-1}]$ \\
\noalign{\smallskip}	
\hline
\noalign{\smallskip}

57169.709	& -12.869	& 0.016\\
57174.706	& -13.005   	& 0.026\\
57177.701	& -12.942	& 0.017\\
57208.728	& -12.078	& 0.014\\
57210.714	& -12.133	& 0.017\\
57236.601	& -12.676	& 0.021\\
57259.755	& -12.959	& 0.014\\
\hline
\end{tabular}
\end{table}

\subsection{Consistency between optical and IR data}

In order to confirm or reject the companion hypothesis of the RV variations observed in the K giant, we used a set of high precision optical and NIR RVs to search for consistency between the two wavelength domains.

The orbital parameters that characterise a Keplerian orbit have been derived from the optical RVs (see Table \ref{parameters}). Because of the smaller number of data points in NIR than in the optical, we only fitted the Keplerian orbit to the GIANO RVs for the RV semi-amplitude ($K_{IR}$) and systemic velocity ($\gamma_{IR}$), keeping the other parameters (P, $T_{P}$, e) fixed at the values derived from the optical RVs. The obtained values in NIR are $\gamma_{IR}= -12,560.42 \pm 6.46$ m/s and $K_{IR}=477.6 \pm 9.5 $ m/s. The different value between center of mass RV ($\gamma$) in optical and NIR enables us to determine the offset between the two datasets. It is important to note that before the observations with GIANO, we have simulated several possible strategies in order to have the best phase coverage and number of observations. We obtained the best results concentrating the observations at phases close to the minimum and maximum, with few observations at the other phases.

\begin{figure}[t]
\centering
\includegraphics[width=\hsize]{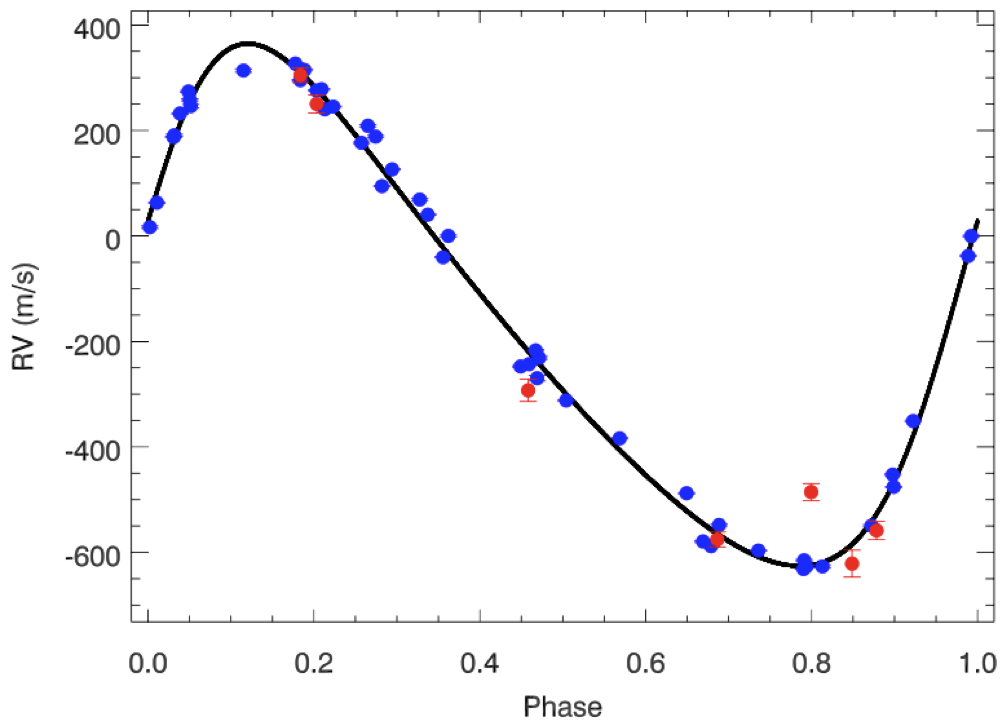}
\includegraphics[width=\hsize]{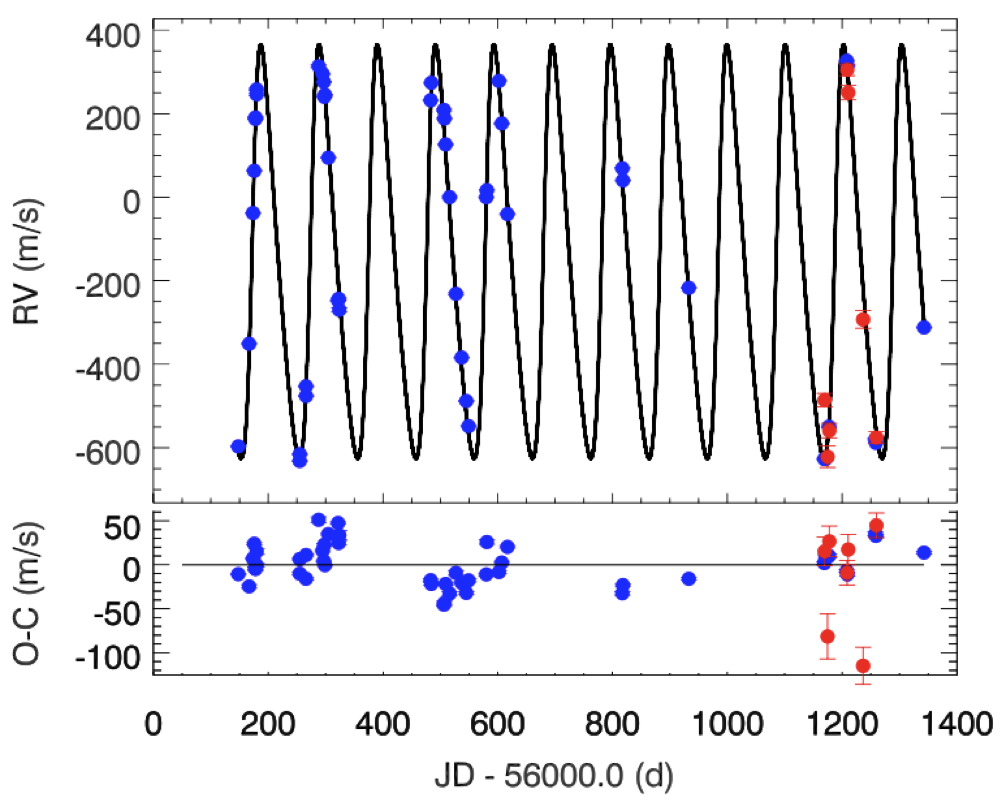}
\caption { \scriptsize  GIANO NIR radial velocity with red circles and HARPS-N optical RVs with blue circles. The best fit Keplerian model obtained from optical data is plotted with a solid black line. \textbf{The top panel} shows the RV data in function of the phase. \textbf{The bottom panel} shows the RV as a function of the time and the residuals information. The NIR radial velocities (in red) are in excellent agreement with the optical data (in blue).}
\label{KG7_phase}
\end{figure}

The seven NIR Doppler points from GIANO plotted in Fig. \ref{KG7_phase} were corrected by offset and median of RV, but were not used for fitting. In Fig. \ref{KG7_phase} we find that the GIANO data follow the Keplerian model predicted by HARPS-N RVs. The NIR data have an excellent phase coverage (one full period) and have consistent amplitude (1.5 $\sigma$) with the optical data. Our results support the hypothesis of an orbiting companion around the K giant star TYC 4282-605-1 with the orbital parameters list on Table \ref{parameters}.

\section{Summary and conclusions}
\label{Summary and conclusions}

We have reported the discovery of a sub-stellar companion around K giant star TYC 4282-605-1. We note that this is the first time that IR+optical RV measurements were used to confirm a planet discovery around a K giant. The main orbital properties are the following: $M_P\sin i = 10.78\pm 0.12 \; M_{\rm J}$; $P=101.54 \pm 0.05$ days; $e = 0.28 \pm 0.01$; $a=0.422 \pm 0.009$ AU.

It is possible to establish an upper-limit to the mass of the low-mass companion taking advantage of the knowledge of the eccentricity and imposing that the minimum distance from the star cannot be smaller than the stellar radius (16 $R_{\odot}$). We obtain an upper limit of 40 $M_{\rm J}$ resulting in a range of mass between 10 and 40 $M_{\rm J}$ for our sub-stellar companion.

\begin{figure}[t]
\centering
\includegraphics[width=\hsize]{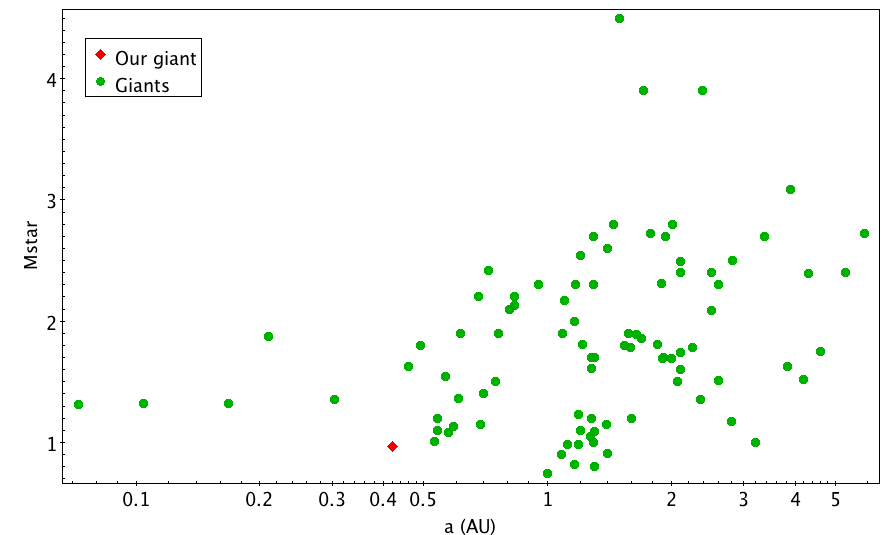}
\caption { \scriptsize  Observed orbital distance vs. the stellar mass for known planets orbiting giant stars taken from Heidelberg group (https://www.lsw.uni-heidelberg.de/users/sreffert/giantplanets/giantplanets.php)}
\label{planets_around_giants}
\end{figure}

Our planet is rather close to its parent star with $asini/R \sim 5$. This poses it close to the limiting value $a/R \sim 3$ that \cite{2014ApJ...794....3V} estimated from the fast tidal decay of the planetary orbits in their models. Indeed they find that only three planets around red giant branch (RGB) stars have $a/R < 10$ considering the observations collected till September 1st 2013. Our star is less massive than most of the RGB hosts that have masses between 1.2 and 2.8 M$_{\odot}$ \citep[cf. Fig. 12 in][]{2014ApJ...794....3V} so the radius inflation during its ascent along the red giant branch was slower giving a longer remaining lifetime to the planet. The orbit of our planet is significantly eccentric and allows us to apply the theory by \cite{1995A&A...296..709V} to estimate the decrease of the eccentricity due to tides inside the red giant host after it left the main sequence. Considering Eqs. (6) and (9) of \cite{1995A&A...296..709V}, we estimate that $e \sim 0.35$ when the star ended its main-sequence evolution because tides did not have time to circularise and shrink the orbit given that the star was still on the initial phase of the red giant branch ascent. If our star were a red clump giant or were ascending the asymptotic giant branch, that is, if it were a core or shell helium-burning object that previously reached the red giant tip, the planet would have been engulfed during the phase of rapid radius increase during the red giant branch ascent and would not be there. We expect that the planet will be engulfed by the star in the next evolutionary phase and indeed today it is one of the closest planet to its host star \citep[see Fig. \ref{planets_around_giants}, see also Fig. 12 in][]{2014ApJ...794....3V}. In other words, the presence of the planet and its eccentric orbit are a strong argument in favour of our previous estimate of the evolutionary phase and mass of our star.

Our goal was to determine the origin of the periodic variability found with HARPS-N in optical band using also NIR radial velocity measurements. The comparison between the amplitude variations in two different bands allowed us to understand the origin of the periodicity of RV time series. 

We obtained NIR RVs with the GIANO spectrograph derived by an ensemble of IDL procedures created to measure RVs with the CCF method. The NIR RVs are in excellent agreement with the optical data obtained with HARPS-N, following the best Keplerian model prediction and being consistent in amplitude. Thus, we confirm the presence of a substellar companion around the giant star after a careful analysis to exclude a stellar origin of the signal thanks to the study of the photometric light curve, the bisector time series and the RV analysis in NIR with GIANO.

In this work we have also studied the chemical abundance of the K giant to explain the nature of the enhancement found for some of $\alpha$-elements. Based on the study done by \cite{2010A&A...513A..35A} and the derived kinematic properties for the target, we can conclude that our K giant it is compatible with a star located in the local thick disk.

We have shown the capability of GIANO to discern the presence of a companion around a giant star and further studies would benefit from the ongoing efforts to use HARPS-N and GIANO-B in a simultaneous way. Thanks to the GIARPS project (GIANO-B + HARPS-N) it is possible to take in a single exposure a high resolution spectrum from 0.383 nm to 2.45 nm. This upgrade will make future observations more efficient, providing precious information on the origin of the radial velocity variations.


\begin{acknowledgements}

This work was supported by WOW from INAF through the \textit{Progetti Premiali} funding scheme of the Italian Ministry of Education, University, and Research.
The research leading to these results has received funding from the European Union Seventh Framework Programme (FP7/2007-2013) under Grant Agree- ment No. 313014 (ETAEARTH).

We thank the referee Artie Hatzes for his useful comments.
      
\end{acknowledgements}

%
%

\bibliographystyle{aa} 
\bibliography{KG7_bibliography.bib} 

\newpage

\begin{appendix} 
\section{Tables}

\begin{table}[h]
\centering
\caption{Measurements of radial velocity with HARPS-N}
\label{Harps_rv_measurments}
\begin{tabular}{cccc}
\hline
\hline
\noalign{\smallskip}
 Epoch [JD]   & RV $\rm [kms^{-1}]$ & $\rm error \, [\rm kms^{-1}]$ \\
\noalign{\smallskip}	
\hline
\noalign{\smallskip}

  56147.839 & 	-12.784 & 	    0.001\\
  56166.789 & 	-12.538 & 		0.002\\
  56173.522 & 	-12.225 & 		0.001\\
  56175.743 & 	-12.124 & 		0.001\\
  56177.793 & 	-12.000 & 		0.001\\
  56177.875 & 	-11.996 & 		0.002\\
  56179.774 & 	-11.929 & 		0.002\\
  56179.888 & 	-11.941 & 		0.003\\
  56254.898 & 	-12.819 & 		0.001\\
  56254.972 & 	-12.802 & 		0.001\\
  56265.822 & 	-12.640 & 		0.001\\
  56265.943 & 	-12.663 & 		0.001\\
  56287.877 & 	-11.873 & 		0.003\\
  56294.842 & 	-11.892 & 		0.001\\
  56296.838 & 	-11.911 & 		0.001\\
  56297.838 & 	-11.947 & 		0.001\\
  56298.862 & 	-11.942 & 		0.001\\
  56304.837 & 	-12.093 & 		0.001\\
  56321.813 & 	-12.435 & 		0.001\\
  56322.841 & 	-12.430 & 		0.002\\
  56323.836 & 	-12.457 & 		0.005\\
  56483.194 & 	-11.955 & 		0.001\\
  56484.233 & 	-11.913 & 		0.001\\
  56506.194 & 	-11.978 & 		0.002\\
  56507.145 & 	-11.999 & 		0.002\\
  56509.159 & 	-12.061 & 		0.001\\
  56516.042 & 	-12.187 & 		0.002\\
  56527.155 & 	-12.418 & 		0.003\\
  56537.019 & 	-12.571 & 		0.001\\
  56545.210 & 	-12.675 & 		0.001\\
  56549.192 & 	-12.735 & 		0.002\\
  56580.046 & 	-12.187 & 		0.001\\
  56581.044 & 	-12.170 & 		0.003\\
  56602.066 & 	-11.909 & 		0.002\\
  56606.945 & 	-12.011 & 		0.001\\
  56616.947 & 	-12.228 & 		0.001\\
  56817.168 & 	-12.118 & 		0.002\\
  56818.143 & 	-12.147 & 		0.001\\
  56932.895 & 	-12.404 & 		0.002\\
  57169.103 & 	-12.814 & 		0.002\\
  57171.085 & 	-12.814 & 		0.003\\
  57177.087 & 	-12.737 & 		0.002\\
  57208.095 & 	-11.860 & 		0.002\\
  57209.211 & 	-11.872 & 		0.002\\
  57258.008 & 	-12.766 & 		0.002\\
  57258.998 & 	-12.775 & 		0.001\\
  57342.797 & 	-12.499 & 		0.001\\
  57350.926 & 	-12.607 & 		0.005\\
\hline
\end{tabular}
\end{table}

\section{Reducing GIANO spectra}

A typical 2D output from a GIANO integration with the previous fibers configuration, is shown in Fig. \ref{2D_output_GIANO}. The whole frame contains 49 groups of 4 arc-shaped tracks where each group represents a spectral order. Any group is composed of two pairs of arcs, each pair corresponding to the output of one of the two optical fibres used to transfer the signal from the telescope focal plane to the spectrograph. Within each order, wavelength increases from right to left. In addition, wavelength decreases with increasing order. One of the goals of the reduction procedure is to find the curve that best fits the signal peak in each sub-track and to count the signal as a function of position within apertures centred on these curves (called traces). In order to obtain an accurate extraction we need a set of calibration frames (dark frames, flat-field frames and reference lamp spectra) and follow the following steps:

\begin{figure}
\centering
\includegraphics[width=\hsize]{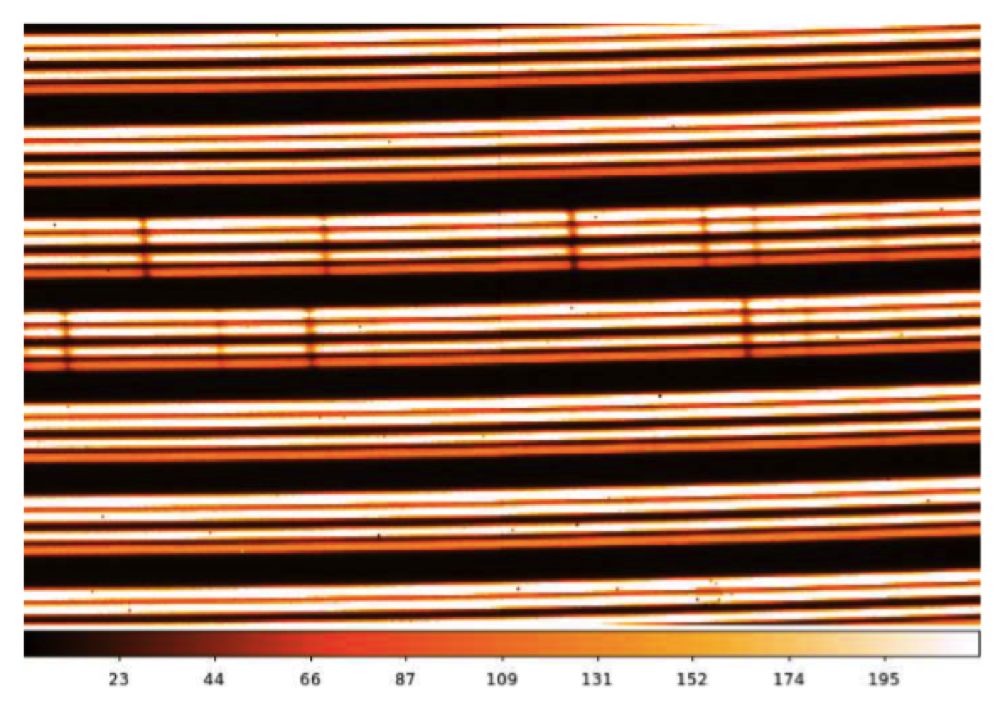}
\caption { \scriptsize  Zoom-in on a Giano image of a continuum lamp. Note that each order is split in four parallel tracks due to the insertion of an image slider that divides the image of the two fibers. }
\label{2D_output_GIANO}
\end{figure}

\begin{itemize}

\item Dark and frame subtraction\\

Dark subtraction is only needed by flat-fields and lamp reference spectra. The dark frames to subtract, must have been taken with the same integration time as the image to be corrected. We have to combine all the dark with the same integration time and then subtract the final image from all the calibration files (flats and UNe).

It is important to remember that the images taken in cycle AB do not need the dark subtraction but it is necessary to remove any residual scattered and diffuse light contribution. For these reasons we subtract frame B from frame A for every target taken in cycle AB and the resulting 2D spectra will exhibit the target signal in all sub-tracks and have most of the bias removed. If there are more than one subtracted AB pair of frames from the same target we have to combine them together before extracting the spectra.

Then, it is necessary to apply bad-pixel corrections to all flat-field frames, calibration-lamp frames and A-B frames. The correction can be done by using the GIANO TOOLS task called CLEAN-UP.\\

\item Flat-field correction\\

The spectrograph efficiency varies with the spectral order and produce differences in peak counts between the four tracks of each order due to the different efficiencies of the two fibres and the positioning of the source in the fibre.

CCDs and NIR detectors exhibit pixel-to-pixel variations in their quantum efficiency. These need being corrected in order to increase the signal-to-noise ratio. The issue is more complicated for images of echelle spectra, due to the large fraction of the detector area (between orders) exposed to a very low flux. To avoid various degrading effects on the noise statistics, flat-field corrections should only be made inside each aperture.

There are two possibles ways of flat-field spectra correction:

\begin{itemize}
\item Construct a 2D flat-field map by deriving a flat-field correction only inside the apertures and divide each frame by this map.

\item Extract the 1D spectrum from each aperture of the flat-field image, normalise it, and divide all extracted (1D) unflatted spectra by the corresponding normalised flat-field 1D spectra.
\end{itemize}

Here we described the 2D technique because making a 2D flat-field frame allows us to define the traces for all orders and check that the subsequent spectrum extraction works. Both methods are equivalent and the same results are obtained.

First of all, we have to combine together all flat-field frames to enhance the signal-to-noise ratio. This is the image we will use to construct the 2D flat-field map and also to find and fit the traces.\\

\item Find traces\\

Following the previous step a final average flat-field image that has the highest signal-to-noise ratio is obtained. Then we fit polynomials to the signal tracks because these polynomials define the traces associated to each track. 

An intermediate step is necessary in order to remove the scattered component because a high level of scattered light can drive small spurious peaks between the central orders exhibiting more counts than the peaks at the highest order. Using the task GIANO FIND TRACE it is possible to construct the scattered light map necessary as a reference file when we will determine the traces.\\

\item Fit traces and construct a 2D flat-field frame\\

It is necessary to derive a polynomial curve fitting the corresponding signal across the detector by using APFLATTEN. It determine the traces that will be used to extract the spectra from any frame.

This IRAF task allows us to check the trace and spectrum fits when we follow the procedure and it gives a final image corresponding with a normalised flat frame.

Then we have to correct all the images (lamp spectra and science frame) dividing (pixel by pixel) each frame by the normalised flat.\\

\item Cloning frame\\

There are 196 traces and 49 orders on the detection (four traces for each order). In fact, for each order there is a bottom trace, a mid-lower one, a mid-upper one and a topmost one. Each trace must be extract and calibrate independently to obtain four 1D for each detection. For this reason we need to make four copies of each detection. The task to use is $copy file$ of GIANO TOOLS. We do the same for the lamp, obtaining four sets of frames for the four different traces.\\

\item Spectrum extraction\\

The task used for extracting spectra is DOECSLIT. However, we first need to run APEDIT in order to associate the traces already found in the flat-field frame to science frame through GIANO FIND TRACE and APFLATTEN.

Before to run APEDIT, we have to split the traces found and re-arrange them in the correct order. To this purpose, we use the GIANO TOOLS task SPLIT FILE operating on the non normalised flat-field frame. The output of this task are four sets of frames for the flat (lower, middown, midup and topmost). 

The next step is to run APEDIT and check that the aperture locations have been set correctly for each frame. We must run APEDIT for all the four track group (lowest, middown, midup and topmost), so that every frame will have its trace set associated.

The next step is DOECSLIT to extract and wavelength calibrate the 1D spectra. This task also has to be done four times, for the four groups of frames. The output file is the extracted spectra in 1D.\\

\item Combining the extracted 1D spectra together\\

At this point, we have four different 1D spectra for each target. It is necessary to combine them together to increase the S/N. The output of each fibre is split into two spots, both feeding the slit. This means that the exposure time is exactly that of the single A or B frame after summing together the pair of 1D spectra corresponding to the same fibre. Consequently, if we add together all four 1D spectra extracted from an AB subtracted frame, the effective exposure time will be twice that of the single A or B frame. If you average together the four 1D spectra, the exposure time will be half that of the single A or B frame.

\end{itemize}

\end{appendix}

\end{document}